\pdfoutput=1

\documentclass{appolb}

\usepackage[varg]{txfonts}   

\usepackage{float} 

\usepackage{graphicx}
 \usepackage{xcolor}

\usepackage{lineno}

\usepackage{subfigure}
\usepackage{bm}
\usepackage[percent]{overpic}
\usepackage{color}

\usepackage{hyperref}
\usepackage[all]{hypcap}
\hypersetup{
    colorlinks,
    citecolor=blue,
    filecolor=blue,
    linkcolor=black,
    urlcolor=blue
}

\def\pt{p_{\rm T}}
\def\av#1{\langle #1 \rangle}
\def 	\r2{\rho_2}

\def\vf{\varphi}

\def\sNN{\mbox{$\sqrt{s_{_{\rm NN}}}$}}

\begin{document}
\title{On resonance contribution to balance functions 
\thanks{Presented at XXV Cracow EPIPHANY Conference.}%
}
\author{Igor Altsybeev
\address{ Saint-Petersburg State University \\ Universitetskaya nab. 7/9, St. Petersburg, 199034, Russia  }
\\
\textit{i.altsybeev@spbu.ru}
}
\maketitle
\begin{abstract}

It is known that resonance decays influence the shape of the charge-balance functions
measured in hadronic collisions.
That is reflected in their rapidity and azimuthal widths and the integral,
and therefore has consequences for different model interpretations.  
In this paper,
we show that the contribution from neutral resonance decays can be removed from
 the balance function in an analytical way, 
and test the performance of the removal procedure with PYTHIA events.
Prospects for applications of the procedure to real data analysis of balance functions
are also discussed.

\end{abstract}
  
\section{Introduction}

Particle production in hadronic collisions
is governed by conservation laws, such as conservation of the local charge.
The charge balance function (BF) has been proposed as a convenient measure of the resulting correlations between the opposite charges in the momentum space~\cite{Bass_2000}.
It is defined as 
\begin{equation}
\label{eq_BF}
B(\Delta y, \Delta \vf)
= {1\over 2} \bigg( \frac{\r2^{+ -}-\r2^{+ +}} {\rho_1^{+}}   +   \frac{\r2^{- +}-\r2^{- -}} {\rho_1^{-}}\bigg) ,
\end{equation}
where $\rho_1^{a}$ is a single-particle density,
$\r2^{a b}$ represents  the density 
of particle pairs 
($a,b$=$+,-$),
and $\Delta y$ and $\Delta\vf$ are differences between two particles in rapidity and in azimuthal angle,
respectively.

Balance functions are typically characterized by their shape, widths in $\Delta y$ and $\Delta\vf$,
and by the integral that is the 
total probability to find the balancing charge within an experimental acceptance.
The width of the balance function in rapidity could 
indicate the time of production of the pair of opposite charges
and provide information about their subsequent transport in the hadronic medium \cite{Bass_2000, 2018_Kapusta_Plumberg, 2012_Pratt},
being affected, however,
by multiple effects like radial flow  \cite{2004_Bozek_BF_in_phi}
quantum statistics \cite{2003_Pratt_Cheng},
etc.

Balance functions for charged particles 
have been measured in Au-Au collisions at RHIC by STAR \cite{2010_STAR_BF}
and at the LHC energies 
in Pb-Pb, p-Pb and pp collisions 
by ALICE \cite{2015_ALICE_BF_pp_pPb_PbPb}.
STAR measured balance functions also for the identified particles (charged kaons and pions) \cite{2010_STAR_BF}.
Recently, preliminary results for BF of kaons and pions in different colliding systems
were presented  by ALICE   \cite{BF_Jinjin_QM18}.
The common observation is that the pionic BF  becomes  narrower in $\Delta y$ and $\Delta\vf$ 
with centrality of A-A collisions,
and it is usually advocated 
to be an indication that hadronization occurs only at the very late stage of the development of the
system.

Influence of neutral resonance decays 
on the shape and the integral of the balance function was discussed, for instance,
in \cite{Bozek_et_al_BF_vs_STAR_2005}.
It was found that the decays of the neutral resonances give about a half of the pion pairs in the rapidity window considered.
In many papers,  however, 
model interpretations of the balance functions 
are given without paying enough attention to 
the role of resonance  decays
and without quantitative estimation of their impact.

In this article, it is demonstrated how the contribution from neutral resonance
decays can be analytically removed from the balance function in order to reveal the  underlying BF shape.
Prospects of application of the resonance removal procedure in real-data analysis are discussed as well.

\section{Balance functions from different particle sources } 

\begin{figure}[b]
\centering
\begin{overpic}[width=0.495\textwidth, trim={0.4cm 1.8cm 0.3cm 1.2cm},clip]
{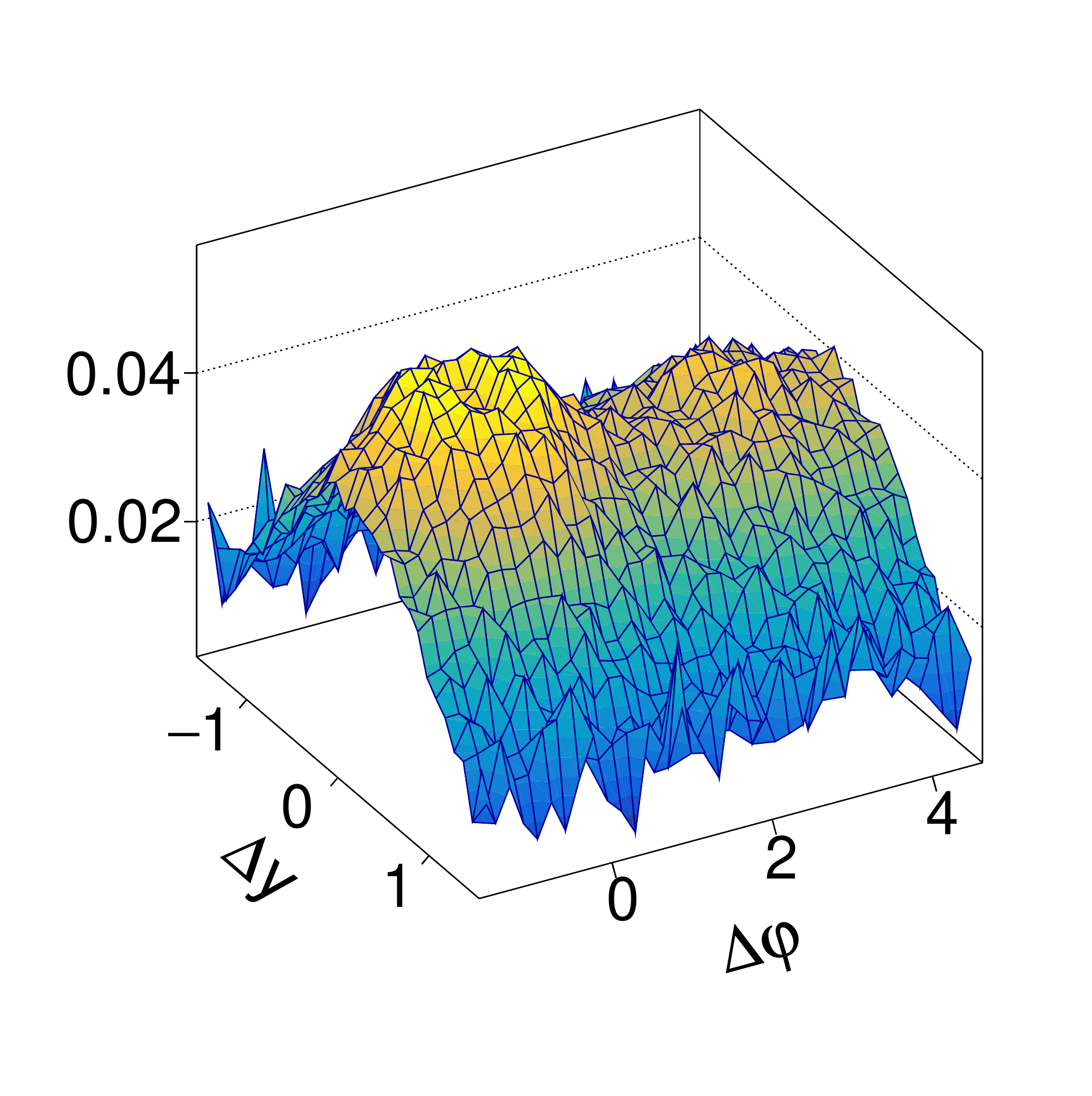}
\put(-4,39){\large \rotatebox{90}{${\rm B}(\Delta y, \Delta\vf) $}}
\put(1,85){ \color{black!65} PYTHIA8 }
\put(1,79){ \color{black!65}   pp@2.76 TeV}
\put(21,68){\large (a) }
\end{overpic}
\hspace{-0.1cm}
\begin{overpic}[width=0.495\textwidth, trim={0.5cm 0.35cm 1.8cm 1.3cm},clip]
{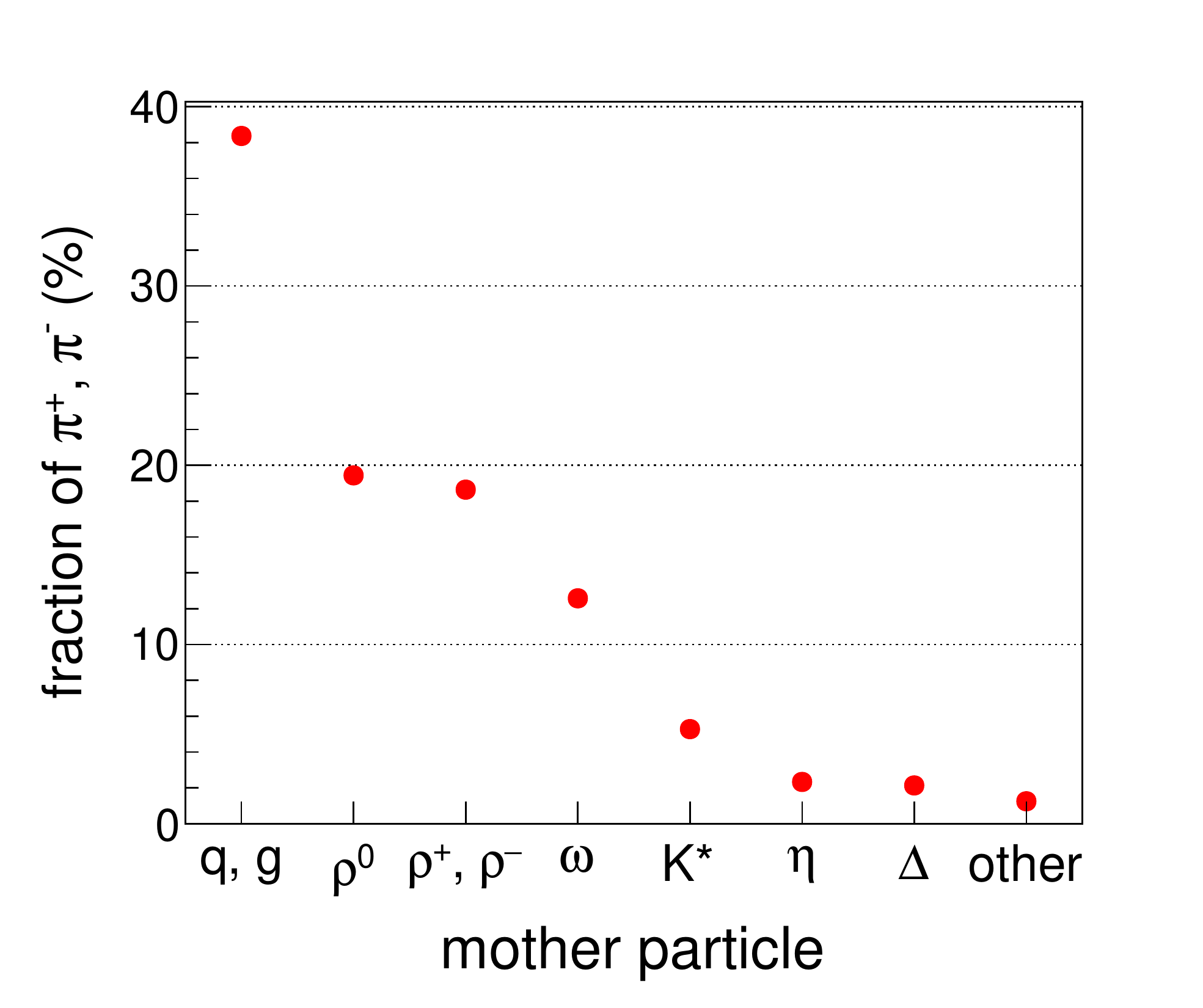}
\put(50,76){\color{black!65} \small PYTHIA8 \small (Monash)}
\put(50,70){\color{black!65} \small pp@2.76 TeV}
\put(52,58){  \small   0.2$<$$\pt$$<$2.0 GeV/$c$}
\put(52,52){  \small $|y|<0.8$ }
\put(20,68){\large (b) }
\end{overpic}
\caption{
(a) Two-dimensional balance function  in PYTHIA8 for pions within ALICE-like acceptance.
\newline
(b) Fractions of charged pions within chosen kinematic cuts
coming from different sources 
in PYTHIA8. 
}
\label{fig:BF_and_pion_sources_in_PYTHIA8} 
\end{figure}
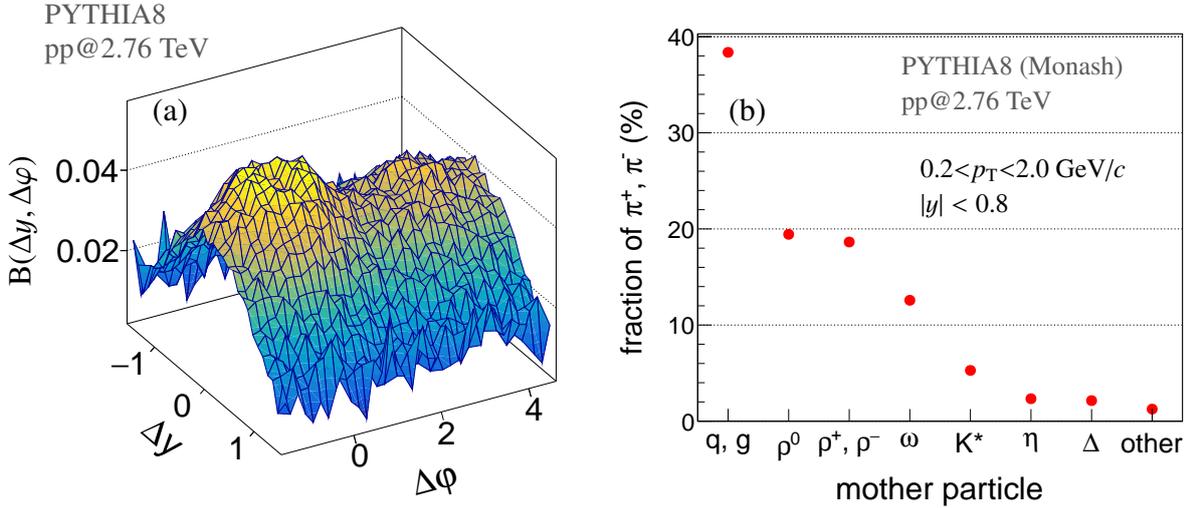

Let us first consider charged-pion balance functions 
from different sources 
in PYTHIA8 \cite{PYTHIA8} in pp collisions at $\sqrt{s}=2.76$ TeV.
ALICE-like kinematic cuts  $|y|<0.8$ and $0.2<\pt<2.0$ GeV/$c$ are adopted for pions.
The two-dimensional BF  shown in Figure \ref{fig:BF_and_pion_sources_in_PYTHIA8} (a) 
demonstrates a typical  near-side peak as well as a broad ridge-like structure along $\Delta\vf$,
which is visible in pp collisions, but 
decreases in more central A-A collisions
(see experimental 2D-plots for BF measured in ALICE in pp, p-Pb and Pb-Pb collisions \cite{BF_Jinjin_QM18_talk}).
However, there is no characteristic ``dip'' at $(\Delta y, \Delta\vf)\approx (0, 0)$ in PYTHIA,
since this dip is attributed to Bose-Einstein correlations that are not present in the generator events by default.

\begin{figure}[t]
\centering
\begin{overpic}[width=0.88\textwidth]{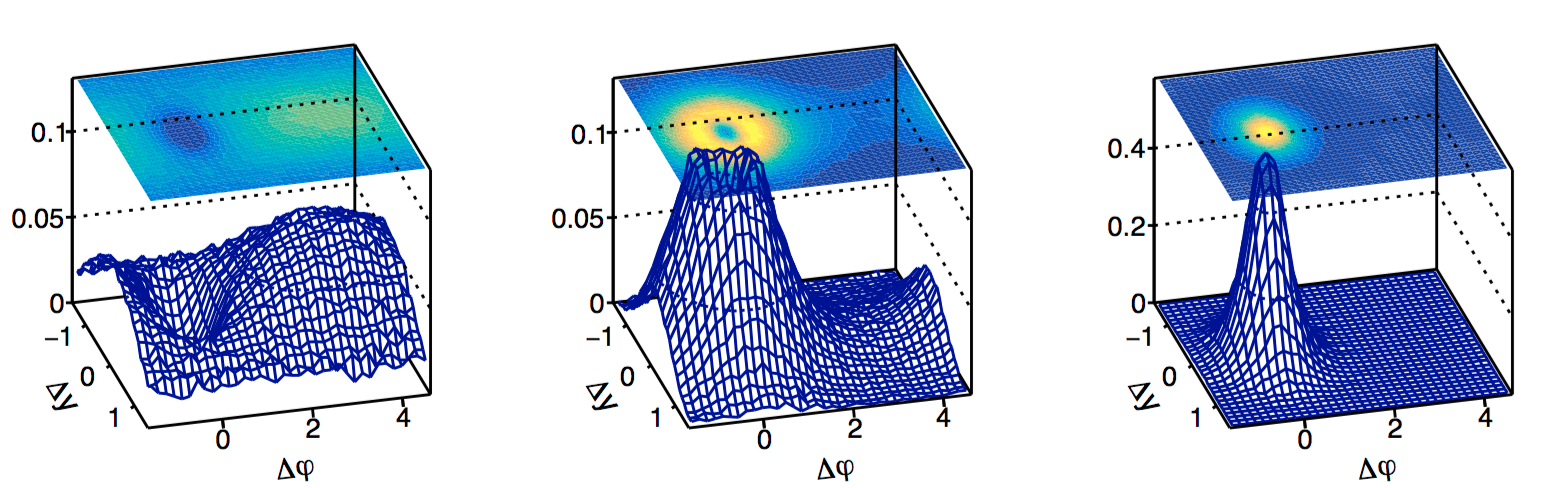}
\put(-3,14){\footnotesize \rotatebox{90}{${\rm B}(\Delta y, \Delta\vf) $}}
\put(10,30){(a) \large   $\rho^0 $}
\put(45,30){(b) \large   $\omega$}
\put(80,30){(c) \large   $\eta$}
\end{overpic}
\caption{
Balance functions of pions 
coming solely from (a) $\rho^0$, (b) $\omega$ and (c) $\eta$ decays
(PYTHIA8, pions within ALICE-like acceptance).
}
\label{fig:BF_2D_rho_omega_eta} 
\end{figure}

Relative abundances of charged pions originating from different sources
are shown in Figure \ref{fig:BF_and_pion_sources_in_PYTHIA8} (b). 
Within the chosen acceptance, 
about 38\% of pions originate directly from quarks or gluons (whatever this means in PYTHIA),
while the rest of pions come from resonance decays.
In particular,
$\approx 35\%$ of all charged pions come from decays 
of {\it neutral} resonances $\rho^0$, $\omega$ and $\eta$.
Corresponding balance functions 
from their decays 
 are shown in
Figure \ref{fig:BF_2D_rho_omega_eta}.
In each case,
the shape of the function is determined solely by the decay kinematics,
which  in most cases corresponds to two-body decays into $\pi^+\pi^-$ 
(in case of $\omega$  the main channel is 
a three-body decay
 into $\pi^+\pi^-\pi^0$, with $\pi^0$ typically being invisible in analysis).
Near the $(0,0)$ point the BF for $\rho^0$ and $\omega$ demonstrate a characteristic volcano-like structure, 
which in case of $\eta$-decays is not visible  
due to the chosen binning of the histogram.

\begin{figure}[b]
\centering
\begin{overpic}[width=0.94\textwidth]
{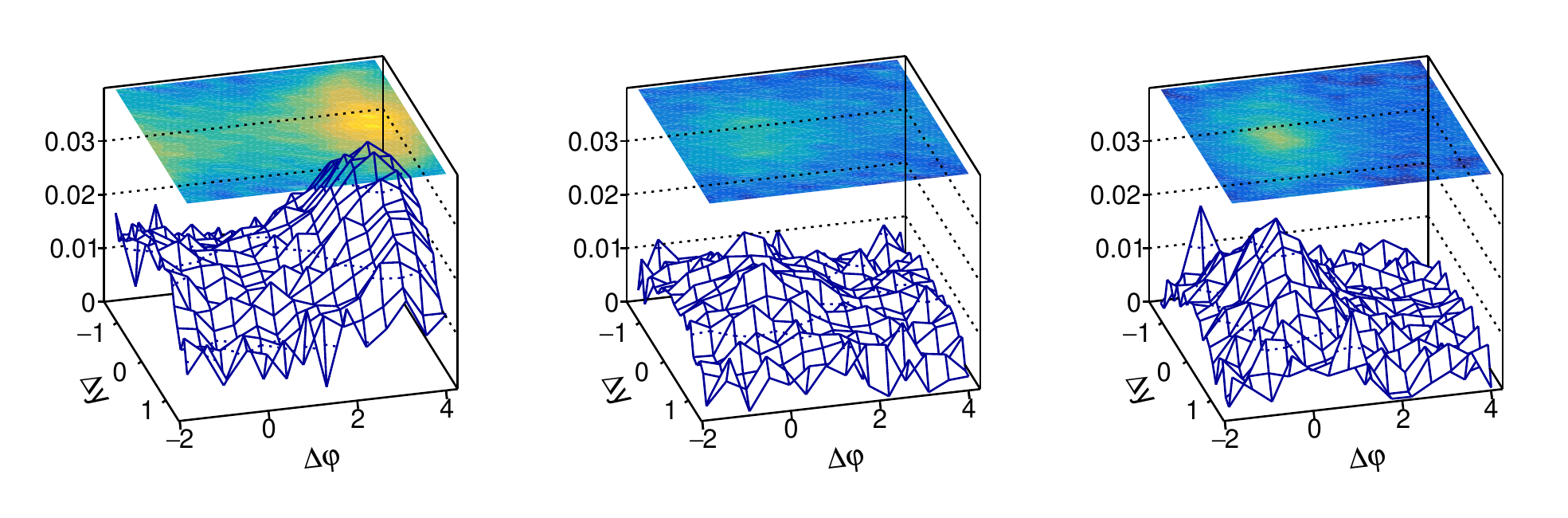}
\put(-1,15){\footnotesize \rotatebox{90}{${\rm B}(\Delta y, \Delta \vf) $}}
\put(2,31){ (a) \small \it quarks and gluons}
\put(41,31){ (b) \large $\rho^+, \rho^-$}
\put(78,31){ (c) \large \it $\Delta$\small-s}
\end{overpic}
\caption{
Pionic balance functions from other sources:
(a) quarks and gluons,
(b)  $\rho^+$ and $\rho^-$,
(c)$\Delta$ hyperons
(PYTHIA8, pions within ALICE-like acceptance).
}
\label{fig:BF_2D_qg_etc} 
\end{figure}
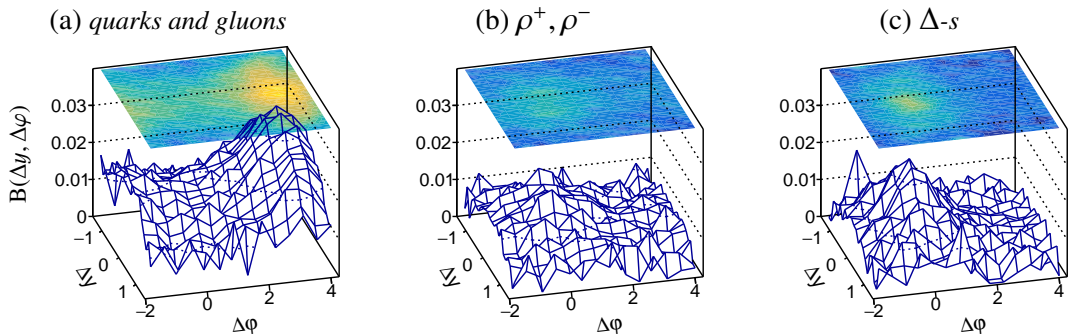

Balance functions for other sources of pions in PYTHIA are shown in Figure~\ref{fig:BF_2D_qg_etc}.
BF for the case when only ``primordial'' pions from quarks and gluons are selected 
for analysis, is shown in panel (a).
The function demonstrates the absence of the near-side structure. 
Instead, there is actually a $y$-broadened structure at $\Delta\vf=\pi$, 
which indicates back-to-back  correlation between the opposite charges,
possibly due to fragmentation of the quark-gluon strings.

Panel (b) of Fig.\ref{fig:BF_2D_qg_etc} shows  BF for pions which originate exclusively from $\rho^+$ and $\rho^-$ decays, neglecting other pions  in  events.
In this case, 
 there are no direct decay-induced correlations between charged pions
--
the correlation  is possible only via 
intrinsic correlations between $\rho^+$ and $\rho^-$ themselves.
As a consequence, 
the shape of the BF is significantly flatter than for the case of neutral resonance decays,  indicating much weaker correlation between $\pi^+$ and $\pi^-$.
Similar conclusion is valid for panel (c), where
 pionic BF is constructed exclusively  from decay products of all types of $\Delta$-hyperons.

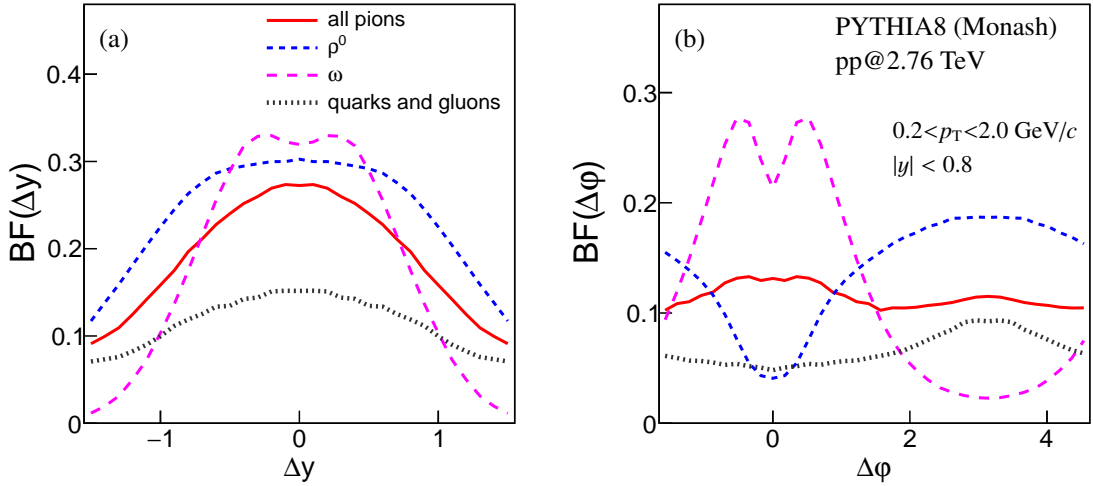
\begin{figure}[H]
\centering 
\begin{overpic}[width=0.99\textwidth]
{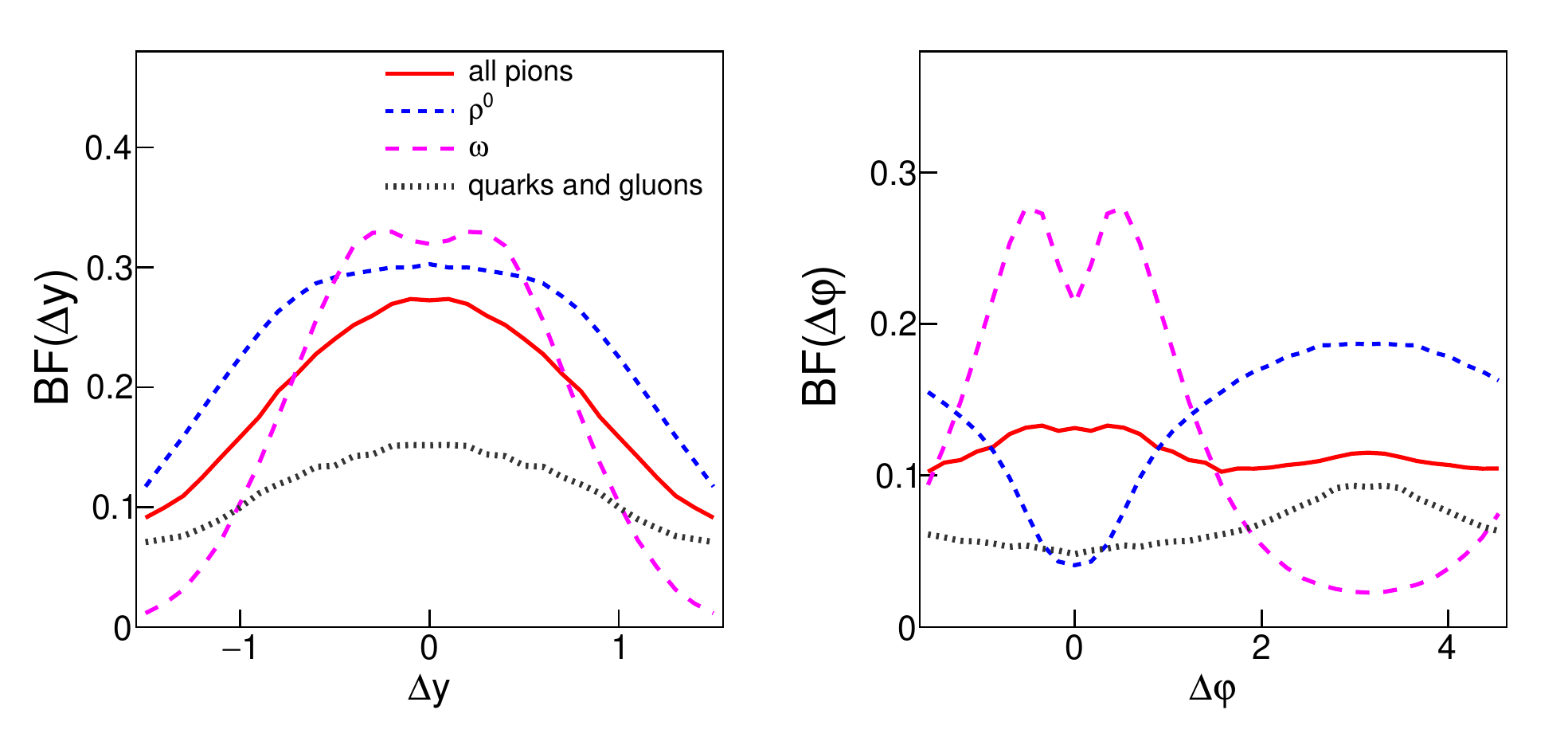}
\put(10,41){\small (a) }
\put(60,41){\small (b) }
\put(74,42){\small PYTHIA8 \small (Monash)}
\put(74,39){\small pp@2.76 TeV}
\put(79,33){\footnotesize   0.2$<$$\pt$$<$2.0 GeV/$c$}
\put(79,30){\footnotesize $|y|<0.8$ }
\end{overpic}
\caption{
Projections of the  balance functions
on (a) $\Delta y$ and (b) $\Delta \varphi$:
red solid lines -- BF measured for all pions in events,
blue dashed lines -- for pions from $\rho^0$ decays,
magenta dashed lines --  for pions from $\omega$ decays.
Gray dotted lines  -- BF for pions from quarks and gluons.
}
\label{fig:BF_PYTHIA_Projections_allPions_res_qg} 
\end{figure}

Fig. \ref{fig:BF_PYTHIA_Projections_allPions_res_qg}. shows
projections on $\Delta y$ and $\Delta\vf$
of all-pion BF (solid line) as well as exclusive balance functions for several pion sources.
It can be seen that the shapes of the functions are significantly different, 
especially in the $\Delta\vf$ projection, where 
BF from $\rho^0$ demonstrates
 a strong depletion at $\Delta\vf$ around zero, while 
in case of $\omega$ decays there is a  peak.
 BF for ``primordial'' pions from quarks and gluons has a ``bump'' at $\Delta\vf \sim \pi$ mentioned above.

\section{Removal of the neutral resonance contributions from BF}

By measuring   the balance functions in the experiment, 
we would like to get insight into
the mechanisms 
of  opposite-charge pair production,
their transport and diffusion in quark-gluon medium, time of hadronisation and so on.
It is known also, that  decays of resonances in final state have a strong contribution to the BF,
and therefore in some sense they distort the signals we desire to measure.
It turns out, however, that 
 it is possible, in principle, to ``purify'' the BF from impact of the {\it neutral} resonance decays,
the corresponding procedure is the subject of this section.

In the paper by Bialas \cite{Bialas_2004}
it was noted that in a model, where particles are produced 
by decays of neutral clusters (these clusters can be correlated), 
contribution to the balance function
from pairs from different clusters cancels in the balance function,
and thus only $(+,-)$ pairs from one cluster do contribute.
This fact was used, for example, to estimate neutral resonance contribution to the BF in Au-Au collisions with STAR data \cite{Bozek_et_al_BF_vs_STAR_2005},
and it was found that the BF shapes of  $\Delta y$ projections for resonant and non-resonant contributions
are nearly the same.

Below we show how to get rid of the resonance contribution in BF analytically.
Let us write a master equation that allows to construct the BF 
for a system of neutral sources. 
Denote number of source  types as $M_s$,
average number of sources of $i$-th type  per event as $\av{k^i}$,
single- and two-particle densities of source decay products as
$\rho_1^i$  and $\r2^i$,
then the balance function can be expressed as

\begin{equation}
\label{master_eq}
{\rm BF}
= \frac{1} {2} 
\frac{
\sum_{i=1}^{M_s} 
\av{k^i}\cdot
\big(
{\r2^i}^{(+,-)} + {\r2^i}^{(-,+)} 
- {\r2^i}^{(+,+)}   -   {\r2^i}^{(-,-)} 
\big)
}
{\sum_{i=1}^{M_s} \av{k^i} \cdot {\rho_1^i}^{(+)}  }.
\end{equation}
With (\ref{master_eq}),
we can 
explicitly remove resonance contributions 
from the BF measured in an experiment,
by subtracting unwanted neutral-source contributions
 from numerator and denominator.
 For that, we need to know single- and two-particle densities for products of a resonance decay,
and single-particle distributions of resonances themselves.
This information is usually available, since experiments do measure
resonance yields and spectra, while the $\rho_1$ and $\rho_2$ of the decay products 
are  determined  by the decay kinematics. 

In order to test the resonance removal procedure, 
PYTHIA events are utilized. 
In each event,
we 
define the four types of neutral sources of charged pions: 
three neutral resonances $\rho^0$, $\omega$ and $\eta$ (contribution to the BF from them
we would like to eliminate),
 and the fourth source is the rest of charged pions -- let us call it a 
``bulk''\footnote{The ``bulk'' can be considered as a neutral object, 
since at the LHC energies $\av{N^+} \approx \av{N^-}$ at mid-rapidity.
The bulk consists of pions from  other resonances
as well as those 
from quark-gluon strings, minijets, etc. 
(see the right panel of the Fig. \ref{fig:BF_and_pion_sources_in_PYTHIA8}).
}.
Note, that the  procedure of resonance removal
can be applied directly to the two-dimensional BF($\Delta y, \Delta\varphi$).
Below, for clarity of representation, only projections of the BF are discussed. 

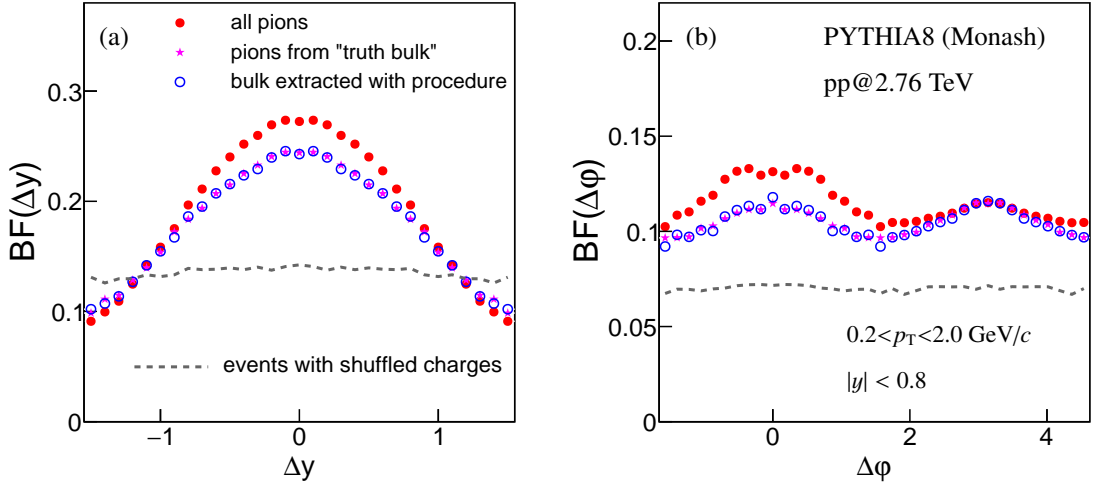
\begin{figure}[H]
\centering
\begin{overpic}[width=0.99\textwidth]
{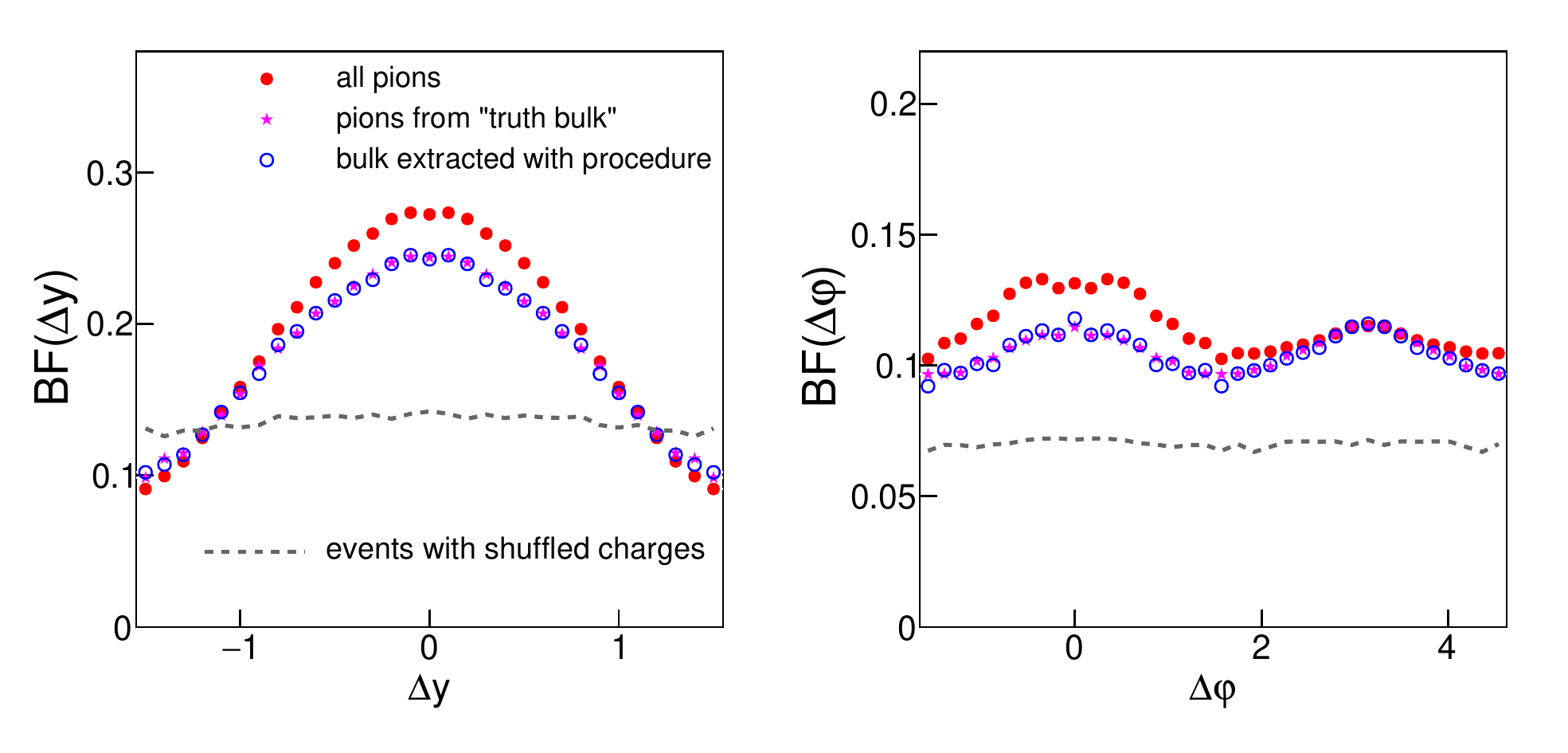}
\put(10,41){\small (a) }
\put(61,41){\small (b) }
\put(73,41){\small PYTHIA8 \small (Monash)}
\put(73,37){\small pp@2.76 TeV}
\put(75,15){\footnotesize   0.2$<$$\pt$$<$2.0 GeV/$c$}
\put(75,11){\footnotesize $|y|<0.8$ }
\end{overpic}
\caption{
Projections of the  balance functions
on (a) $\Delta y$ and (b) $\Delta \varphi$:
closed circles -- BF measured for all pions in events,
star markers -- for pions from the ``bulk" (see definition in the text),
open circles --  BF of the  ``bulk'' extracted with the resonance removal procedure.
Dashed curves -- BF for events with shuffled charges within $|y|<2$.
}
\label{fig:BF_PYTHIA_Projections} 
\end{figure}

Figure \ref{fig:BF_PYTHIA_Projections}
demonstrates projections on $\Delta y$ and $\Delta \vf$
of the initial BF for all pions (closed circles), 
BF of the ``true'' bulk (stars) and BF of the bulk extracted from the initial BF with the resonance removal procedure (open circles).
It can be seen that the points of the extracted bulk
are on top of the ``true'' points.
We may note also that without the $\rho^0$, $\omega$ and $\eta$ resonances 
the balance function in a region $|\Delta y|\lesssim 1$ and $|\Delta \vf| \lesssim 2$ 
 significantly deviates from the initial BF,
which affects the BF widths in $\Delta y$ and $\Delta \vf$
as well as the BF integral (from 0.361 for the all-pion BF to 0.330 for  the ``bulk" BF).

On the same plots, the dashed lines show the BF for the case when electric charges 
of pions in each event are shuffled within $|y|<2$, 
``washing out'' the angular correlations in this range.
Balance functions for shuffled events can be considered as  baselines, 
with which the measured BF should be compared.

Note that, in principle, with the expression~(\ref{master_eq})
we can construct balance functions 
 as combinations of other kinds of charge-neutral sources, 
in particular,  quark-gluon strings can be considered (if they are long enough in rapidity 
so that charges at the string ends do not play a role). 
It would be interesting, for example, to consider centrality dependence of the BF
in the models where several kinds of strings
with varied particle emission functions (which depend, for instance, on string tension parameter)
are packed together \cite{ea_vv_sigma, vk_strongly}, or in the model with repulsive interaction between strings, where $\rho_2$ for each string is modified in a laboratory frame 
by the flow-like effect due to string repulsions
\cite{abramovskii_1998, string_repulsion}.
In such models,
shape of the BF is determined by local charge conservation in string fragmentation process,
and modified further by the decays of resonances which are produced from strings.

\section{Prospects for application of the removal procedure to real-data BF   }

The resonance removal procedure 
can be applied to real data, for instance, to BF measured in pp, p-A and A-A collisions.
For that, as it was mentioned above, 
it is necessary to know yields of resonances 
and single- and two-particle densities
for their  decay products, which evolve with centrality of the collision.

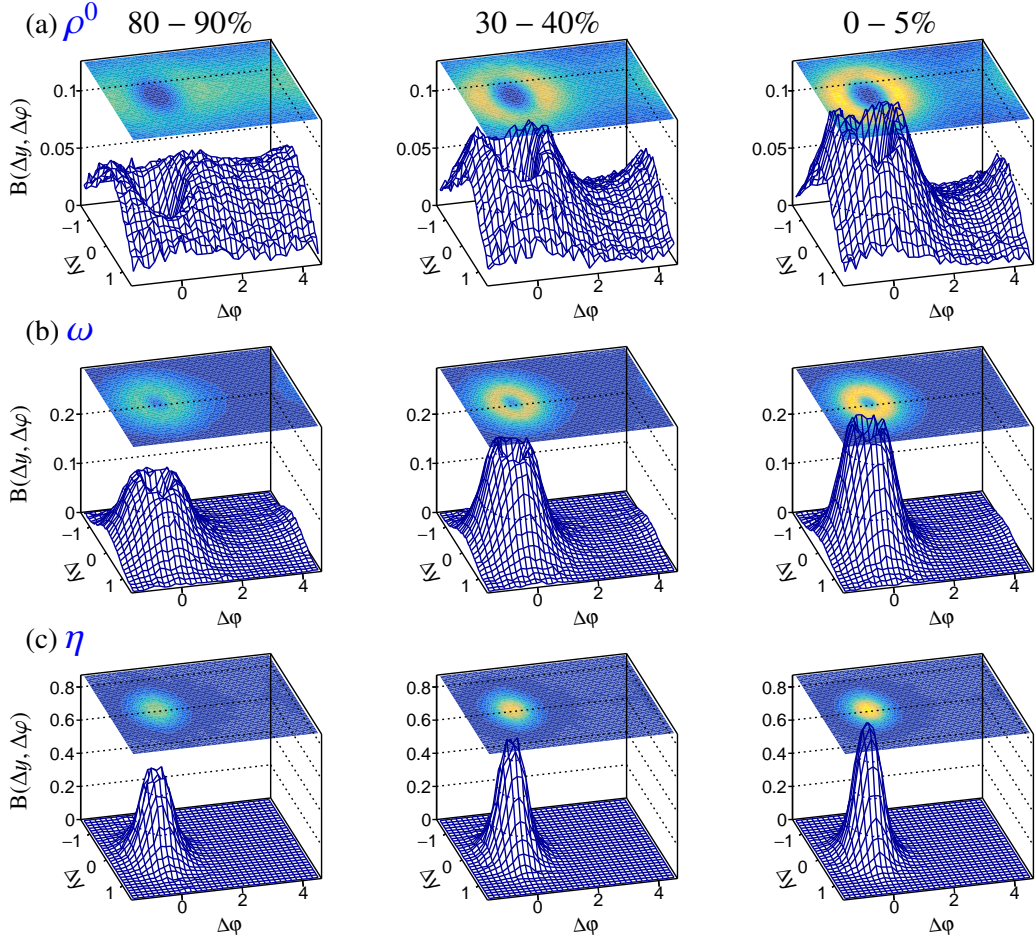
\begin{figure}[b]
\centering
\begin{overpic}[width=0.88\textwidth, trim={0.1cm 0.25cm 0.75cm 0.5cm},clip]
{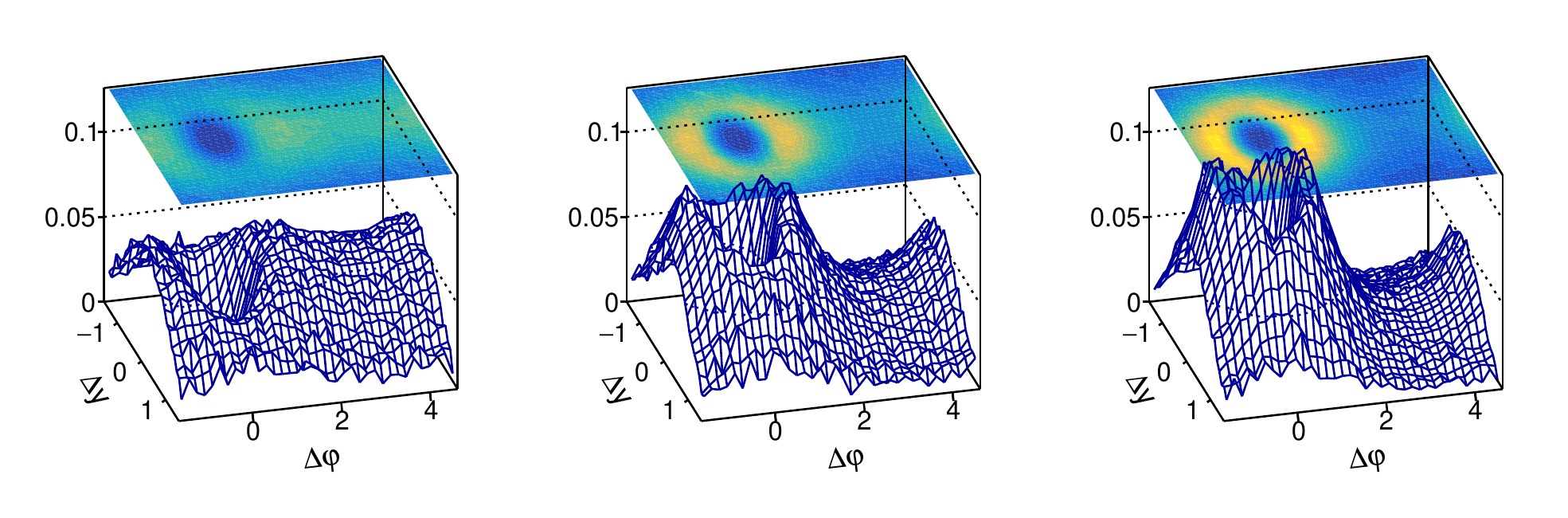}
\put(-0.5,14){\footnotesize \rotatebox{90}{${\rm B}(\Delta y, \Delta\varphi) $}}
\put(1,30){(a) \color{blue} \Large   $\rho^0 $}
\put(11,30){\large $80-90\%$}
\put(45,30){\large $30-40\%$}
\put(81,30){\large $0-5\%$}
\end{overpic}
\begin{overpic}[width=0.88\textwidth, trim={0.1cm 0.25cm 0.75cm 0.5cm},clip]
{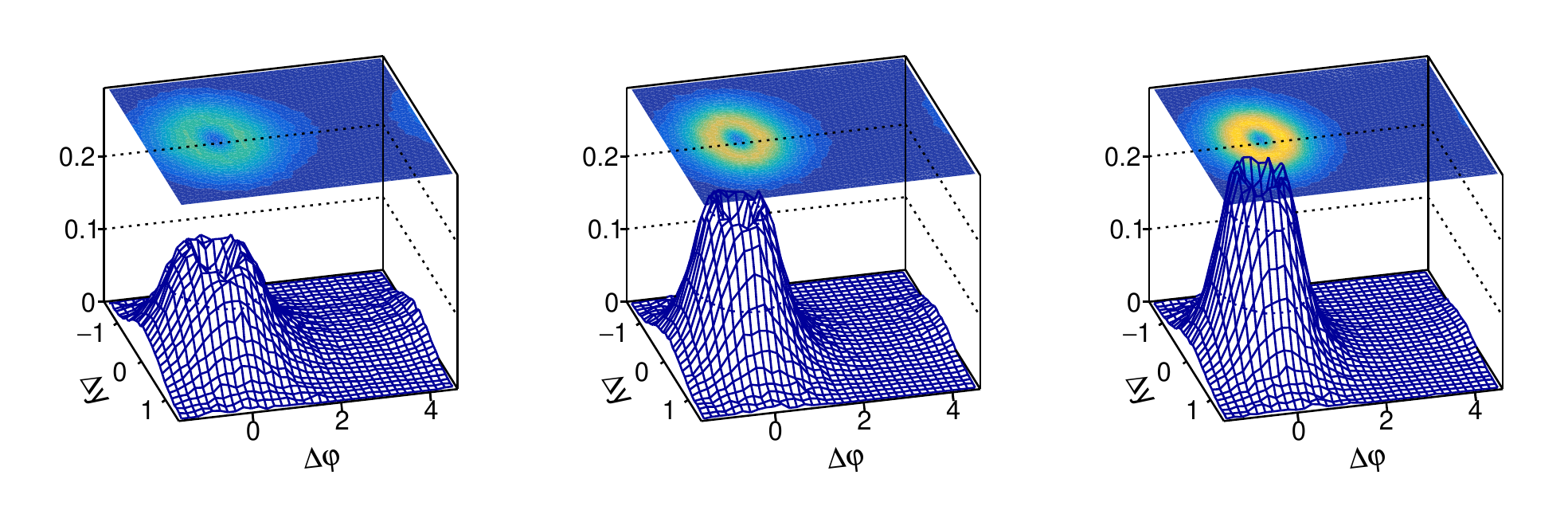}
\put(-0.5,14){\footnotesize \rotatebox{90}{${\rm B}(\Delta y, \Delta\varphi) $}}
\put(1,30){(b) \color{blue} \Large   $\omega$}
\end{overpic}
\begin{overpic}[width=0.88\textwidth, trim={0.1cm 0.25cm 0.75cm 0.5cm},clip]
{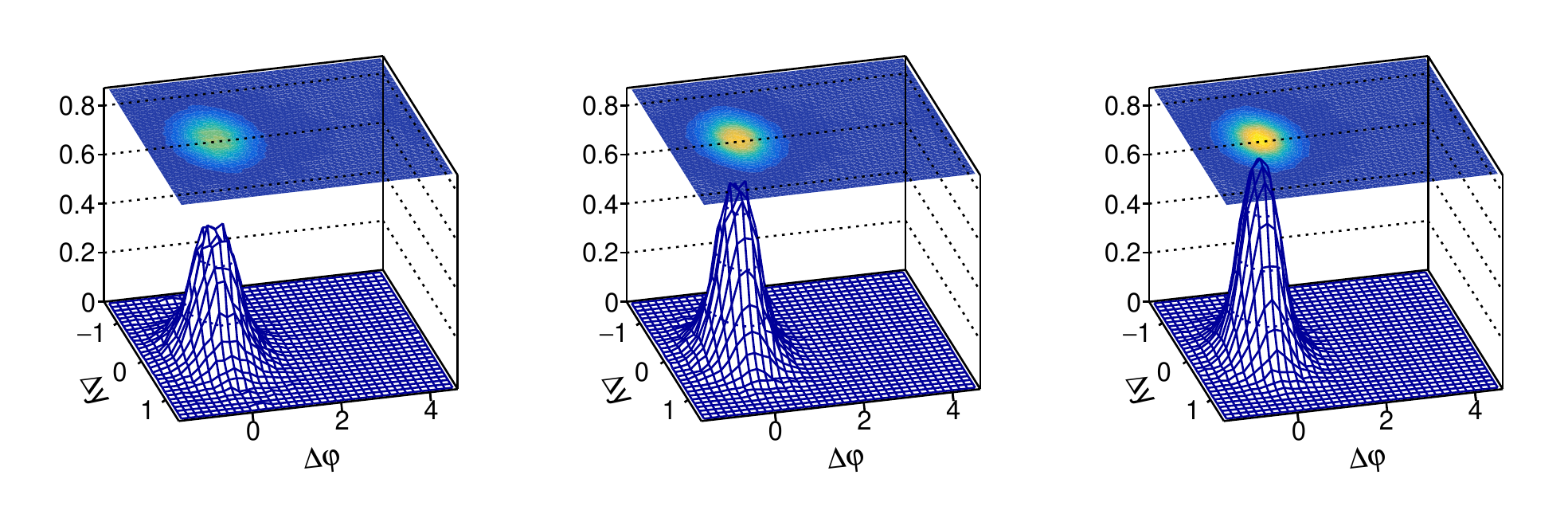}
\put(-0.5,14){\footnotesize \rotatebox{90}{${\rm B}(\Delta y, \Delta\varphi) $}}
\put(1,30){(c) \color{blue} \Large   $\eta$}
\end{overpic}
\caption{
Evolution of the shape of the pionic 2D balance function from
 (a) $\rho^0$, (b) $\omega$ and  (c) $\eta$
 decays
due to hardening of the spectra with centrality of Pb-Pb collisions $\sNN=2.76$ TeV.
Spectra of resonances 
 are obtained by utilizing blast-wave fit parameters from  \cite{ALICE_pi_K_p_PbPb}.
Acceptance for pions is 0.2$<$$\pt$$<$2.0 GeV/$c$,  $|y|<0.8$.
}
\label{fig:BF_blast_wave_evolution} 
\end{figure}

For example, for analysis of pionic balance functions in ALICE data,
one may use
relative fractions of $2\rho^0/(\pi^++\pi^-)$ measured by ALICE \cite{ALICE_rho0},
which are about 0.1, and the yields of $\omega$, that are very similar.
However, the measured transverse momentum ($\pt$) spectra for $\rho^0$ 
at different centralities
are not precise enough to be used for calculation of resonance
pair densities $\rho_2$.
Instead,
one can make approximations for resonance spectra,
for example, 
take  blast wave fit
parameters from  $\pi$, K, p spectra analysis in ALICE \cite{ALICE_pi_K_p_PbPb},
construct spectra of resonances of desired types,
 sample them and apply simple decay kinematics
 in order to obtain necessary densities of the decay products.
For illustration, the balance functions solely from  
(a) $\rho^0$, (b) $\omega$ and  (c) $\eta$
decays are plotted
in Fig.\ref{fig:BF_blast_wave_evolution},
where 
in different columns
spectra 
correspond to peripheral, mid-central and central Pb-Pb collisions. We may note how the shape of the BF changes towards central events -- 
while spectra become harder,
width of the near-side peak narrows,
also, in case of $\rho^0$ the ridge-like structure along $\Delta\vf$ decreases.
The centrality evolution of the BF shown in Fig.\ref{fig:BF_blast_wave_evolution} 
can be qualitatively compared with  two-dimensional 
preliminary plots presented by ALICE in \cite{2015_ALICE_BF_pp_pPb_PbPb}.

The effect of balance function narrowing with hardening of the spectra
is a more general phenomenon, valid not only for resonances.
For instance, in thermal models,
the higher the transverse velocity of particles at the freeze-out surface,
the closer the distance between balancing charges in rapidity and azimuth \cite{2004_Bozek_BF_in_phi}.
In the balance function, an interplay 
between the magnitude of the radial flow of the ``bulk'' and boosted neutral resonances
 may, in principle, be resolved with analytical procedure (\ref{master_eq}).

\begin{figure}[t]
\centering
\begin{overpic}[width=0.33\textwidth, trim={0.2cm 0.8cm 1.1cm 0.7cm},clip]
{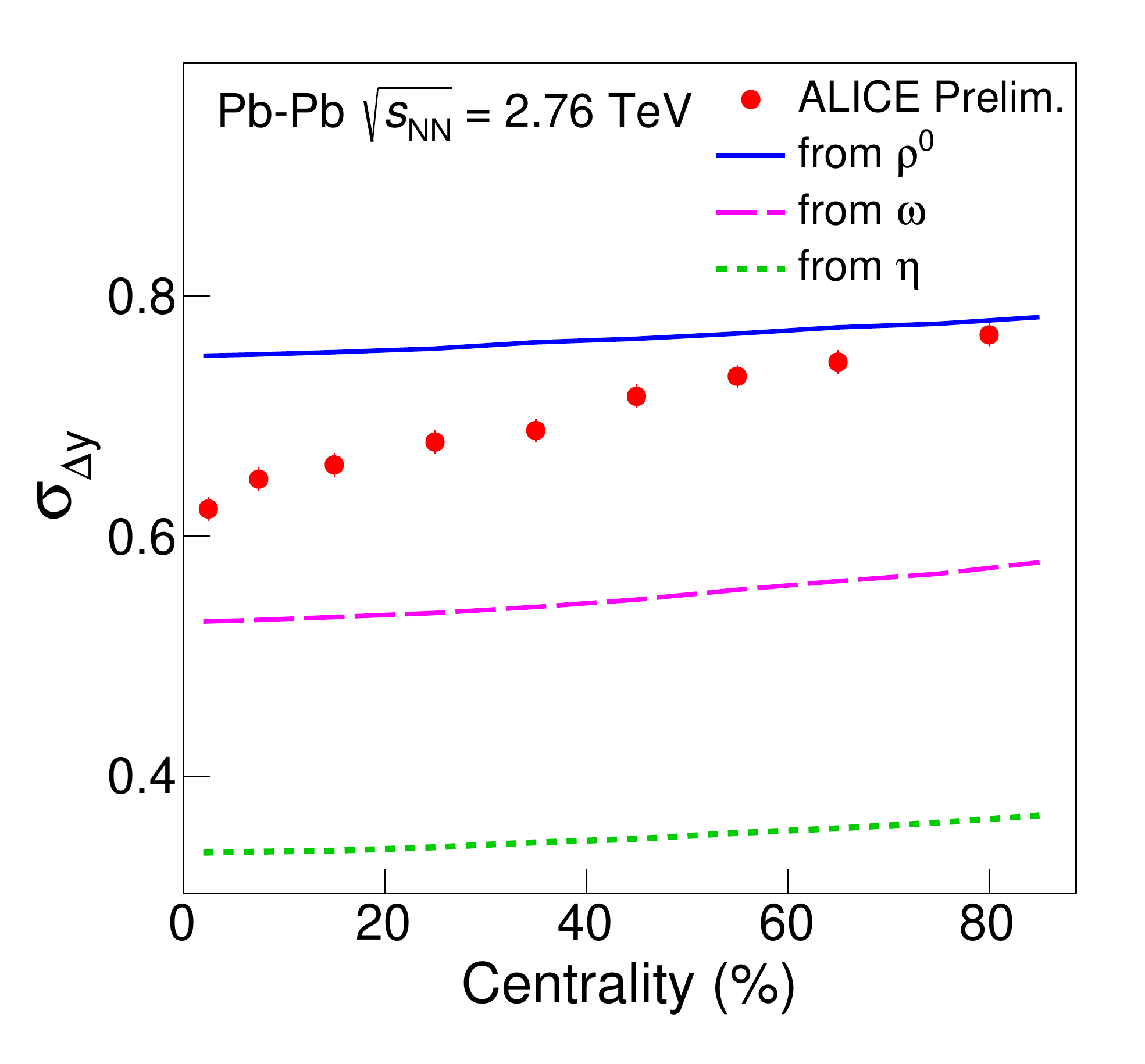}
\put(21,20){\small (a) }
\end{overpic}
\hspace{-0.2cm}
\begin{overpic}[width=0.33\textwidth, trim={0.2cm 0.8cm 1.1cm 0.7cm},clip]
{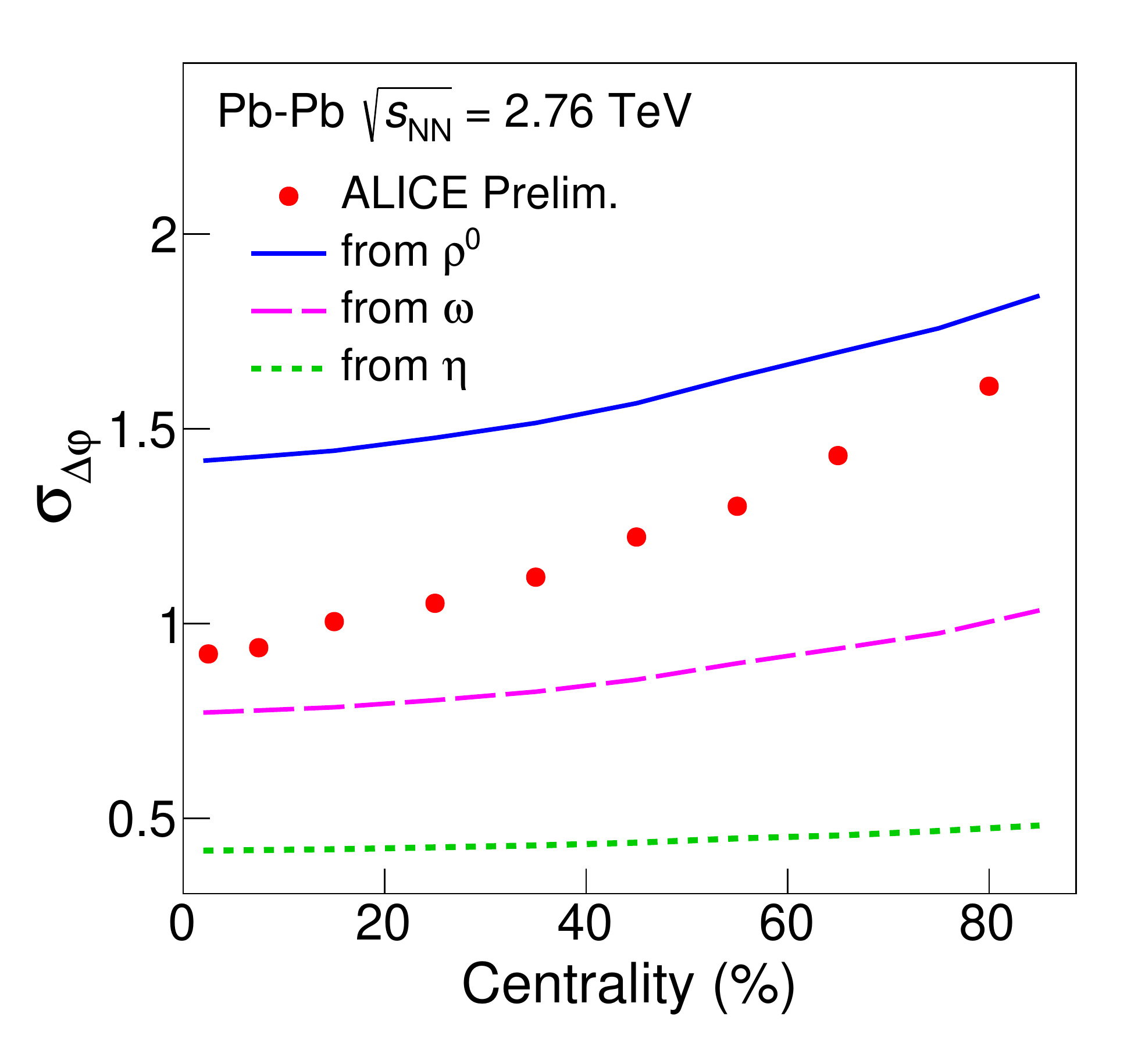}
\put(21,20){\small (b) }
\end{overpic}
\hspace{-0.2cm}
\begin{overpic}[width=0.33\textwidth, trim={0.2cm 0.8cm 1.1cm 0.7cm},clip]
{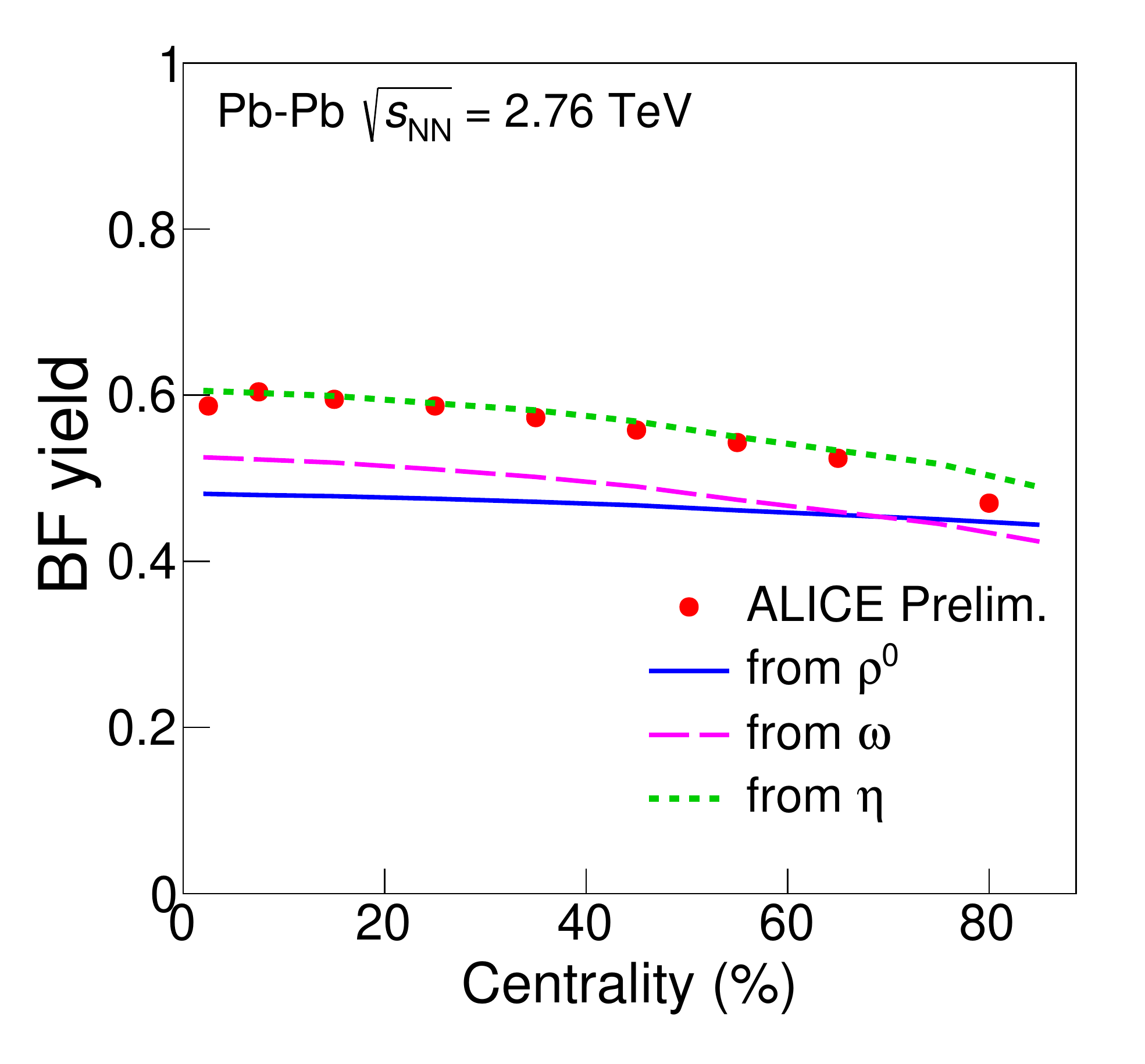}
\put(21,20){\small (c) }
\end{overpic}
\caption{
Widths of the pionic balance functions for $\rho^0$,  $\omega$ and  $\eta$ decays
 in  (a)  $\Delta y$ and (b)~$\Delta \vf$ dimensions
as a function of centrality, 
in comparison to the ALICE preliminary results \cite{BF_Jinjin_QM18}.
Spectra of resonances are obtained by utilizing the blast-wave fit parameters from  \cite{ALICE_pi_K_p_PbPb}.
Panel (c) shows  comparison of the BF integrals over the ALICE-like acceptance.
}
\label{fig:BF_widths_integral_after_blast_wave} 
\end{figure}

Narrowing of the pionic BF widths 
in $\Delta y$  and $\Delta \vf$  projections
for $\rho^0$,  $\omega$ and  $\eta$ decays
with hardening of the spectra towards central events
is demonstrated   in Fig.\ref{fig:BF_widths_integral_after_blast_wave}
(a, b).
Preliminary ALICE results    \cite{BF_Jinjin_QM18}
are plotted as well for comparison.
Of course, the blast-wave approximation for resonance spectra
is quite rude\footnote{Note,
that we can use the same expression (\ref{master_eq}) 
in a more ``differential" way, namely,
to construct 
BF for any given neutral resonance spectrum
by treating resonances within each $\pt$ bin as an independent kind of source.
Similar trick can be done with bins of the resonance mass peak.
}, especially for $\rho^0$, which has a broad mass spectrum
and a lifetime of about $c\tau\approx 1.3$~fm/$c$, so that 
daughter pions can rescatter in the surrounding medium, 
altering the $\rho^0$ spectrum and the structure of the balance function \cite{2018_BF_Pratt}.
In addition, relative fractions of resonances change with centrality \cite{ALICE_rho0},
and an interplay between the resonance yields 
and abundances of ``primordial'' pions
at different centralities may partially be responsible for the observed centrality dependence 
of the real-data BF widths. 

Panel (c) in Fig.\ref{fig:BF_widths_integral_after_blast_wave}
shows BF integrals for resonances at different centralities within ALICE acceptance. 
Larger integrals in central events indicate
 higher probability for each particle to observe an oppositely charged partner within the acceptance,
which is just another indication of narrowing of the  BF  for sources boosted in transverse direction.

Application of the resonance removal procedure 
to real experimental data is out of the scope of the present paper.
As a final remark in this section, 
we note,
that 
in analysis of real experimental data
there could be other undesired neutral-source contributions to the BF.
Namely, 
weak decays $\rm K^0_S\rightarrow \pi^+\pi^-$
may noticeably contribute to the pionic balance function, if track selection cuts are not tight enough
to reject secondary particles from weak decays.
This problem is relevant also for other types of balance functions,
for example, the BF between pions and protons may contain a  contribution
from $\rm \Lambda^0 \rightarrow \pi p$ decays.
Such ``parasitic'' contributions from weak decays
can be eliminated with the neutral source removal procedure (\ref{master_eq}),
provided that the proper simulation of detector response exists 
and fractions of secondary particles are known.

\section{Conclusions}

In this article, we  discussed how 
resonance decays influence the shape of the charge-balance function,
its rapidity and azimuthal widths, the integral,
and indicate that proper treatment of resonance contributions may
 have important consequences for different model interpretations of the BF.
It was shown
how the contribution from neutral resonance decays can be analytically removed from
 the BF measured in an experiment.
The procedure was tested with PYTHIA events. 
As an example, it was demonstrated, that
after removal of contributions from $\rho^0$,  $\omega$ and  $\eta$ decays
the shape of the near-side peak of the pionic BF in $\Delta y$ and $\Delta \vf$  
is visibly modified.

Removal of the neutral resonance contributions from real-data balance functions 
measured in pp, p-A and A-A collisions
is possible  
in case when the resonance yields  
and their spectra are known  in each centrality (multiplicity) class with enough precision.
The procedure may be also of a practical use
 for  purification of  balance functions from contamination by products of weak decays of neutral particles
(like  $\rm K^0_S$ and $\rm \Lambda^0$).
The described procedure can be applied to two-dimensional balance functions, not only for their 1D projections.

\section*{Acknowledgements}
This work is supported by the Russian Science Foundation, grant 17-72-20045.


\end{document}